\documentclass[aps,prl,floatfix,preprint]{revtex4-1}
\usepackage{graphicx}
\usepackage{
	amsmath,amssymb,mathtools,bm,textcase
	}
\usepackage{multirow}
\usepackage{etoolbox}% http://ctan.org/pkg/etoolbox

\newcommand\tablesplit[2]{\begin{tabular}{c}#1\\#2\end{tabular}}

\begin{document}
	\title{
		Conformality of Charge Density Wave
	}
	\author{
		Keiji Nakatsugawa$^{1,3}$, Tatsuhiko N. Ikeda$^{4}$, Takeshi Toshima$^{5}$, and Satoshi Tanda$^{1,2}$}
	\vspace*{.2in}
	\affiliation{
			$^{1}$Center of Education and Research for Topological Science and Technology, Hokkaido University, Sapporo, 060-8628, Japan. 
			\\
			$^{2}$Department of Applied Physics, Hokkaido University, Sapporo, 060-8628, Japan.
			\\ 
			$^{3}$International Center for Materials Nanoarchitectonics, National Institute for Materials Science, Tsukuba, 305-0044, Japan.
			\\
			$^{4}$ Institute for Solid State Physics, University of Tokyo, Kashiwa, 277-8581, Japan
			\\
			$^{5}$ Department of General Education, National Institute of Technology, Toyama College, Toyama, 939-8630, Japan
		}

	\date{\today}
	\vspace*{2ex}
	\begin{abstract}

		New quantum phenomena are continuously being discovered in 2D systems. 
		In particular, the charge density wave (CDW) has the aspect of a quantum crystal with a macroscopic wave function (order parameter), so unlike quantum liquids (superconductivity, quantum Hall liquids $^3$He, $^4$He), new ground states such as supersolid and Moir\'e solids can be expected. 
		However, it is difficult to describe these states because of their quantum aspect, hence there is still no theory that can explain CDW phases in a unified way.
		The best way to describe a quantum crystal seems to be a conformal transformation that allows local deformation (wave properties) and preserves local angles (crystal properties).
		
		Here, we propose a unifying conformal description of 2D CDW phases in the typical 2D CDW material transition metal dichalcogenides (MX\textsubscript{2}). We discover that the discommensurate CDW phases in MX\textsubscript{2} can be explained beautifully by a discrete conformal transformation of CDW wavevectors. This conformality is due to commensurability of CDW with the MX\textsubscript{2} lattice. In other words, interference of harmonic wavefunction induces conformality. 
		
		Using this new conformal formulation, we explain experimental nearly-commensurate/stripe/T CDW phases in 1\textit{T}-TaS\textsubscript{2} ($\sqrt{13}\times\sqrt{13}$ structure), 2\textit{H}-TaSe\textsubscript{2} ($\sqrt{9}\times\sqrt{9}$ structure), and explain the origin of a new experimental nearly-commensurate phase in TaSe\textsubscript{2} thin-film ($\sqrt{7}\times\sqrt{7}$ structure).
		
		This theory is very simple in the sense that it includes only discommensuration and comprises physics as rich as quantum Hall liquids. This new description will broaden our perspective of quantum crystals.
	\end{abstract}
	\maketitle
	\section{Introduction}
		Commensurability is a major factor which determines the phase of crystalline systems \cite{Bak_1982}. For a one-dimensional system, two superimposed crystals are said to be commensurate (C) if the ratio of their lattice constant is a simple fractional number: Some of the lattice points of the two lattices coincide with a larger period. Conversely, if this ratio is an irrational number, then the two crystals are incommensurate (IC). Commensurability can be extended to two-dimensional crystals, which is defined by the lattice constants and orientation of superimposed crystals.

		Charge density waves (CDWs) are typical systems in which commensurability plays a central role \cite{Wilson_Review_1975, Gruner_Review, Monceau_Review}. These are electronic crystals which are described by macroscopic order parameters like superconductors, superfluids and fractional quantum Hall liquids (FQHLs). Several mechanisms of CDW formation have been proposed \cite{Gruner_Review,vanHove,Rice-Scott,Neto,Monceau_Review,Yang2000} and new CDW phases with domain walls (discommensurations) are continually discovered \cite{Yoshida, Toshima, Hidden}. 
		
		Throughout this article, we focus on transition metal dichalcogenides (MX\textsubscript{2}). MX\textsubscript{2} are layered compounds which induce 2D CDWs with wavevectors $\mathbf Q^{(i)}\ (i=1,2,3)$ satisfying the triple-Q condition $\mathbf Q^{(1)}+\mathbf Q^{(2)}+\mathbf Q^{(3)}=\mathbf 0$. These compounds exhibit distinct CDW phases with different $\mathbf{Q}^{(i)}$, including IC, nearly-commensurate (NC), C, stripe \cite{Fleming}, and T \cite{TPhase}. A phenomenological Ginzburg-Landau model of CDW in MX\textsubscript{2} was first introduced by McMillan \cite{McMillan3} to explain the IC to C phase transition and discommensuration. 	Nakanishi and Shiba extended this free energy theory to explain the appearance of the NC phase in 1$T$-TaS\textsubscript{2} and 2$H$-TaSe\textsubscript{2}  using CDW harmonics \cite{NS1T,NS2H}. In general, the search of CDW free energy local minima is a difficult task. We revisited the Nakanishi-Shiba-McMillan models for monolayer 1$T$-TaS\textsubscript{2} and 2$H$-TaSe\textsubscript{2} and discovered the multivalley free-energy landscape, where free energy local minima correspond to different CDW phases including the stripe and T phases \cite{Multivalley}. The existence of these CDW phases in monolayer MX\textsubscript{2} (pure 2D system) is confirmed both experimentally \cite{Sakabe} and theoretically \cite{Multivalley}. However, the underlying mechanism for the appearance of free energy local minima is not yet understood. 
		
		Therefore, a unifying understanding the CDW phases, and perhaps other 2D phases, including the Moir\'e solids, seems to require a new method. This method must connect the IC (continuous) and C (discrete) phases. On the other hand, a CDW has the aspect of a quantum crystal with a macroscopic wave function (order parameter). So, we surmise that such a method is a conformal transformation that allows local deformation (wave properties) and preserves local angles (crystal properties).

		In this article, we propose a unifying conformal description of 2D CDW phases. We study CDW phases using a unique property of 2D systems, that is, a 2D vector $\mathbf{Q}=(Q_x,Q_y)$ is equivalent to a complex number $Q=Q_x+\mathrm{i}Q_y$ ($\mathrm{i}^2=-1$). We introduce a conformal method with complex numbers and discover that discommensurate (NC, stripe, T) CDW phases are described by distinct complex integers called Eisenstein integers. More specifically,
		we discover that the multivalley CDW free energy structure in MX\textsubscript{2} systems \cite{Multivalley}, where free energy local minima correspond to different CDW phases, can be explained beautifully by a discrete conformal transformation with $\mathbb{Z}[\omega]$. 
		Then, we apply our conformal formulation to explain experimental discommensurate CDW phases in 1\textit{T}-TaS\textsubscript{2} ($\sqrt{13}\times\sqrt{13}$ structure), 2\textit{H}-TaSe\textsubscript{2} ($\sqrt{9}\times\sqrt{9}$ structure), and explain the origin of a new experimental NC phase in TaSe\textsubscript{2} thin-film ($\sqrt{7}\times\sqrt{7}$ structure).

	\section{Description of CDW with Eisenstein integers}
		\subsection{The Commensurability Structure}
			Consider a triangular lattice (MX\textsubscript{2} monolayer) with primitive lattice vectors $\mathbf{a}_1=(1,0)a$ and $\mathbf{a}_2=\frac{1}{2}(-1,\sqrt{3})a$ where $a=\lVert \mathbf{a}_1\rVert=\lVert \mathbf{a}_2\rVert$ and $\lVert \cdot\rVert$ represents the norm of a vector. The CDW superlattice vector $\mathbf{A}_\mathrm{C}$ of the C phase is defined as \cite{Wilson_Review_1975}
			\begin{align}
				\mathbf{A}_\mathrm{C}&=(\mu+\nu)\mathbf{a}_{1}+\nu \mathbf{a}_{2},
			\end{align}
			where $\mu$ and $\nu$ are positive integers that depend on materials (e.g., $\mu=3$ and $\nu=1$ for the $\sqrt{13}\times\sqrt{13}$ structure in 1$T$-TaS\textsubscript{2} as shown in Fig. 1(a)).
			The corresponding reciprocal-lattice vectors are defined as
			\begin{align}
				\mathbf{Q}_\mathrm{C}&=\frac{(\mu+\nu)\mathbf{G}_{1}+\nu\mathbf{G}_{2}}{\lVert\mathbf{A}_\mathrm{C}\rVert^2/a^2}
				\label{QcDef}
			\end{align}
			where $\lVert\mathbf{A}_\mathrm{C}\rVert=\sqrt{\mu^2+\mu\nu+\nu^2}$  and $\mathbf{G}_{i}$ ($i=1,2$) are rotated reciprocal lattice vectors satisfying $\mathbf{a}_{i}\cdot \mathbf{G}_{j}=(3\delta_{ij}-1)\pi$ (see Supplementary Note \ref{SupplG}). There are actually three equivalent superlattice vectors $\mathbf{A}^{(1)}_\mathrm{C}, \mathbf{A}^{(2)}_\mathrm{C}, \mathbf{A}^{(3)}_\mathrm{C}$ which are rotated from $\mathbf{A}_\mathrm{C}$ by $0^\circ$, $120^\circ$, and $240^\circ$, respectively, and three equivalent wavevectors $\mathbf Q^{(i)}_\mathrm{C}$ which are defined likewise (Figure \ref{Fig_Geometry}). For brevity we omit the superscript when unnecessary.

			The above relations between two-dimensional vectors in the real and reciprocal spaces are concisely represented using complex numbers.
			The correspondence $(1,0)\leftrightarrow 1$ and $(0,1)\leftrightarrow \mathrm{i}=\sqrt{-1}$ maps $\mathbf{a}_{1}$ and $\mathbf{a}_{2}$ to $a$ and $a\omega$, respectively, where $\omega={e^{2\pi\mathrm{i}/3}}=-\frac{1}{2}+\frac{\sqrt{3}}{2}\mathrm{i}$. 
			Then, the triangular lattice can be mapped to the set of \emph{Eisenstein integers} $\mathbb{Z}[\omega]$ with elements $z=m+n\omega$ ($m,n\in\mathbb{Z}$) \cite{Conway1996}.  Consequently, $\mathbf{A}^{(i)}_\mathrm{C}$ are equivalent to the Eisenstein integers (Fig. \ref{Fig_Geometry} (a))
			\begin{equation}
				A_\mathrm{C}=[(\mu+\nu)+\nu\omega]a,\quad A^{(i)}_\mathrm{C}=A_\mathrm{C}\omega^{i-1}.
			\end{equation}
			$\mathbf{G}_i$ are mapped to $G_i=2\pi a^{-1} \omega^{1-i}$ and $\mathbf{A}_\mathrm{C}\cdot \mathbf{Q}_\mathrm{C}=2\pi$ corresponds to $A_\mathrm{C}^{\ast}Q_\mathrm{C}=2\pi$ where the asterisk denotes complex conjugation  (this correspondence can be formalized, for instance, using 2D Clifford algebra \cite{Hestenes}). Consequently, $\mathbf{Q}_\mathrm{C}^{(i)}$ are equivalent to
			\begin{equation}
				Q_{\mathrm{C}}=\frac{2\pi}{A_\mathrm{C}^{\ast}}=\frac{G_1}{\mu-\nu\omega},\quad Q_\mathrm{C}^{(i)}=Q_\mathrm{C}\omega^{i-1},
				\label{QComplex}
			\end{equation}
			where we used Eq. (3) and $\omega^* = -1 - \omega$. For brevity we set $G_1=1$ (i.e., $a=2\pi$) in the following.
			%Note that the triple-Q condition $\mathbf Q^{(1)}+\mathbf Q^{(2)}+\mathbf Q^{(3)}=\mathbf 0$ corresponds to the identity $\omega^2+\omega+1=0$ of Eisenstein integers. 

		%
		\subsection{Discommensuration}
			Next, we consider general CDW phases in equilibrium. 
			MX\textsubscript{2} exhibit three types of isotropic CDW phases where each $\mathbf Q^{(i)}$ are separated by $120^\circ$: IC, NC, and C. There are also the anisotropic stripe phase [32] and the T phase [33].
			The IC wavevector $\mathbf{Q}_\mathrm{IC}^{(i)}$ is usually determined by Fermi surface nesting or van Hove singularity \cite{Gruner_Review,vanHove,Rice-Scott,Neto,Monceau_Review,Yang2000} which is independent of the underlying lattice.
			On the other hand, the C wavevector $\mathbf{Q}_\mathrm{C}^{(i)}$ is determined from lattice symmetry and satisfies the commensurability condition $A_\mathrm{C}^{\ast}Q_\mathrm{C}=2\pi$. 
			
\begin{figure}
			\includegraphics[width=\linewidth]{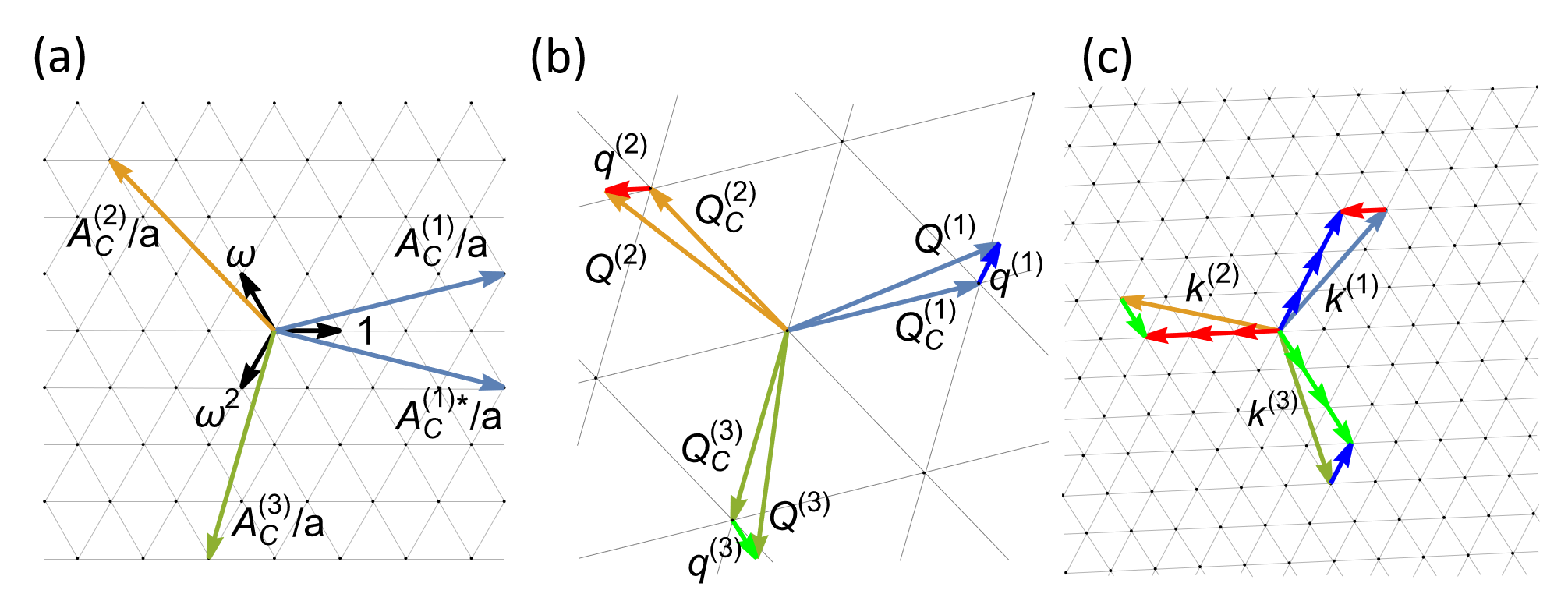}
			\caption{
				%[Now, I can generalize this figure to Q vectors with nonequal norm.]
			(a) C-CDW superlattice vectors can be represented by Eisenstein integers
			%$A^{(i)}/a=(\mu+\nu)+\nu\omega$ and C-CDW wavevectors $Q_\mathrm{C}^{(i)}=2\pi/A_\mathrm{C}^{(i)\ast}$ satisfy the commensurate structure $\mu Q^{(i)}_\mathrm{C}-\nu Q^{(i+1)}_\mathrm{C}=G_i$ 
			(this Figure shows the $\sqrt{13}\times\sqrt{13}$ structure with $\mu=3$ and $\nu=1$
			%and $Q_\mathrm{C}^{(1)}=G_1/(3-\omega)$
			). (b) Domain wall wavevectors $q^{(i)}$ are given as deviation from $Q_\mathrm{C}^{(i)}$. (c) CDW harmonics $k^{(i)}$ satisfy the commensurability condition $\mu q^{(i)}-\nu q^{(i+1)}=k^{(i)}$.}
			\label{Fig_Geometry}
		\end{figure}

			NC phases between the C phase and the IC phase have domain walls (discommensurations) which originate from commensurability with the underlying base lattice \cite{McMillan3,Yamada_Takatera,NS1T,NS2H}. 
			Let us write general CDW wavevectors $\mathbf{Q}^{(i)}=(Q^{(i)}_x,Q^{(i)}_y)$ using complex wavenumbers $Q^{(i)}=Q^{(i)}_x+\mathrm{i}Q^{(i)}_y$.
			Then, the domain wall wavevectors $\mathbf q^{(i)}$ and CDW harmonic wavevectors $\mathbf k^{(i)}$ are also written using complex numbers (Figure \ref{Fig_Geometry})
			\begin{align}
				q^{(i)}&=Q^{(i)}-Q_\mathrm{C}^{(i)}
				\\
				k^{(i)}&=\mu q^{(i)}-\nu q^{(i+1)}
			\end{align}
			{Equipped with these complex numbers, we are ready to give a conformal description of CDW phases.}
			If a CDW did not interact with the underlying base lattice, then $Q^{(i)}$ may take arbitrary continuous values. In real systems, however, commensurability with the base lattice lead to discrete values of $Q^{(i)}$. These discrete values form a conformal group as we show in the next section.

	\section{Conformality of CDW phases}
		Suppose that the IC wavevectors $\mathbf Q_\mathrm{IC}^{(i)}$ and C wavevectors $\mathbf Q_\mathrm{C}^{(i)}$ are given. Experimentally, CDW wavevectors in MX\textsubscript{2} satisfy the triple-Q condition $\mathbf Q^{(1)}+\mathbf Q^{(2)}+\mathbf Q^{(3)}=\mathbf{0}$. Then, the NC phase transition is described by the Nakanishi-Shiba-McMillan phenomenological free energy\cite{McMillan3,NTYS,NS1T,NS2H}
			\begin{align}
				F&=\int d^2r\bigg\{a(\mathbf{r})\alpha(\mathbf{r})^2-b(\mathbf{r})\alpha(\mathbf{r})^3+c(\mathbf{r})\alpha(\mathbf{r})^4
				\label{FreeEnergy}
				\\
				+&d(\mathbf{r})\sum_{i=1}^3|\psi_i(\mathbf{r})\psi_{i+1}(\mathbf{r})|^2
				+\sum_{i=1}^3\psi_i(\mathbf{r})^\ast e_i(-i\mathbf\nabla)\psi_i(\mathbf{r})\bigg\}.\nonumber
			\end{align}
			Here, $\mathbf{r}$ is the 2D spatial coordinate. The complex order parameters $\psi_i(\mathbf{r})$ are related to the charge density $\rho(\mathbf r)=\rho_0(\mathbf{r})[1+\alpha(\mathbf r)]$ where $\alpha(\mathbf r)=\text{Re}\left[\psi_1(\mathbf r)+\psi_2(\mathbf r)+\psi_3(\mathbf r)\right]$. $\rho_0(\mathbf r)$ is the charge density in the normal phase. Each coefficient has periodicity of the base lattice and are written as $a(\mathbf r)=a_0+2a_1\sum_{i=1}^3\cos(\mathbf G_i\cdot\mathbf r)$, etc. Moreover, {$e_i(\mathbf{Q})=s|\mathbf{Q}-\mathbf Q_\text{IC}^{(i)}|^2/|\mathbf G_i|^2$} where $s$ is a constant (see Supplementary Note \ref{SupplType} for consistency with other forms of $e_i(\mathbf{Q}^{(i)})$).
			The free energy \eqref{FreeEnergy} can be integrated using the Nakanishi-Shiba expansion \cite{NS1T,NS2H}
			\begin{equation}
				\psi_i(\mathbf r)=\sum_{\substack{l,m,n\geq0\\l\cdot m\cdot n=0}}\Delta_{lmn}^{(i)}\exp\{\mathrm{i} \mathbf Q_{lmn}^{(i)}\cdot\mathbf r\},
				\label{philmn}
			\end{equation}
			where $l,m,n$ are integers, $\mathbf Q_{lmn}^{(i)}=\mathbf Q^{(i)}+l\mathbf k^{(i)}+m\mathbf k^{(i+1)}+n\mathbf k^{(i+2)}$, and $\mathbf{k}^{(i)}$ are CDW harmonics defined as in the previous section. 
			
			In previous studies of CDW free energy, the the free energy local minima are obtained by calculating the free energy $F$ for all values of $\mathbf{Q}$. However, we observe that local minima can also be obtained by differentiating $F$ with respect to $\mathbf{Q}$. This major simplification can be formalized as follows. First, after integration of $F$, the wavevector dependence appears only in the terms proportional to $e_i$. Second, in our previous work, we reported the free-energy landscape where different free-energy local minima correspond to distinct $l,m,n$ values (Ref. \cite{Multivalley} and Supplementary Figure 2). Therefore, we can treat each $\Delta_{lmn}^{(i)}$ as different degrees of freedom. Then, we observe that the CDW wavevectors between the C and IC phases (free energy local minima) are given as solutions of
			\begin{align}
				\frac{\partial^2 F}{\partial\Delta^{(i)}_{lmn}\partial \mathbf{Q}^{(i)}}
				\propto\Delta_{lmn}^{(i)\ast}
				\frac{\partial e_i(\mathbf{Q}_{lmn}^{(i)})}{\partial\mathbf{Q}^{(i)}}=\mathbf{0}.
				\label{MinEqn}
			\end{align}
			{This equation can be solved readily and we obtain $\mathbf Q_{lmn}^{(i)}=\mathbf Q_\text{IC}^{(i)}$ as conditions for local minima (Supplementary Note \ref{SupplMinEqn}). In the case of isotropic NC phases, we have the following simplification: (i) the summation in Eq. \eqref{philmn} is a summation over Eisenstein integers $z_{lmn}=l+m\omega+n\omega^2$ (note the identity $\omega^2+\omega+1=0$); (ii) the CDW harmonics have the simple expression $k=Q/Q_\mathrm{C}-1$; the wavevectors are separated by $120^\circ$, which implies $Q^{(i+1)}=\omega Q^{(i)}$. Then, we obtain}
			\begin{equation*}
				Q_\mathrm{NC}+z_{lmn}[Q_\mathrm{NC}(\mu-\nu\omega)-1]=Q_\mathrm{IC}.
			\end{equation*}
			
\begin{figure}[th]
				\begin{center}
					\includegraphics[width=0.65\linewidth]{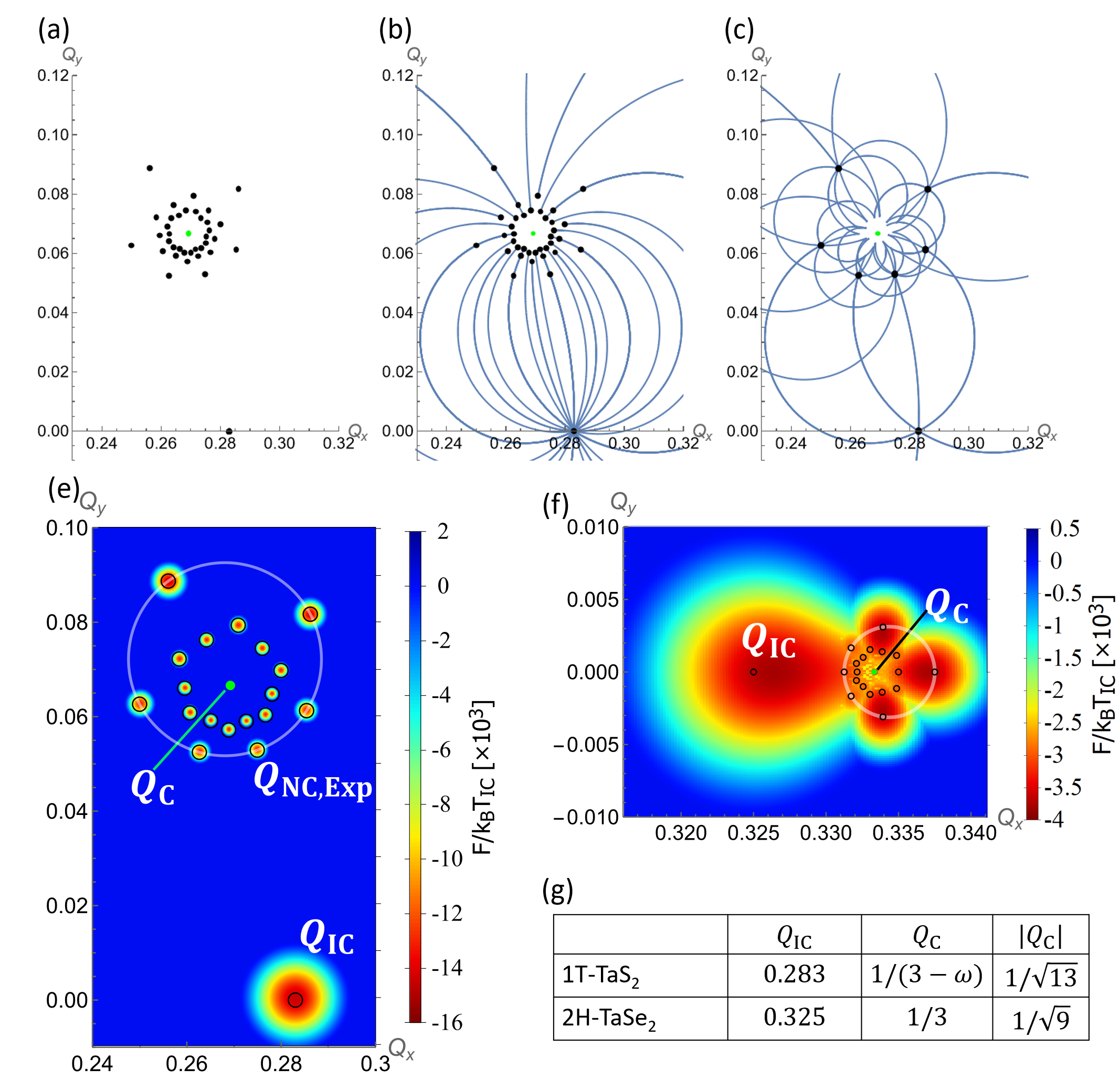}
				\end{center}
				\caption{Different CDW free energy local minima (CDW phases) correspond to distinct Eisenstein integers $z_{lmn}=l+m\omega+n\omega^2$ (in units with $G_1=1$). The green dot represents the C phase and the black dots (circles in (e) and (f)) represent local minima. The blue circles show Eq. \eqref{LocalMinimaModular} but regarding $l,m,n$ as continuous variables (that is, they describes the flow of group action). 
				(a) CDW local minima with $\mu=3$ and $\nu=1$ ($\sqrt{13}\times\sqrt{13}$ commensurate structure) is shown for $N=3$ harmonics (see also Supplementary Video 1). 
				(b) $f_{lmn}(Q_\mathrm{IC})$ for $|z_{lmn}|\leq 3$. This type of transformation can explain how each local minima are obtained from the IC phase. 
				(c) $f_{lmn}(f_{l'm'n'}(Q_\mathrm{IC}))$ for $|z_{lmn}|,|z_{l'm'n'}|\leq 1$. This transformation can explain successive CDW phase transitions. 
				The free energy landscape for (e) $1T-$TaS\textsubscript{2} and (f) $2H-$TaSe\textsubscript{2} are obtained as in our previous work \cite{Multivalley} (except for $2H$-TaSe\textsubscript{2} we used $T-T_\mathrm{IC}=-0.1$ to show local minima more clearly). The free-energy calculation agrees with the local minima of Eq. \eqref{LocalMinimaModular}. 
				We used experimental values of wavevectors in (g).
				}
				\label{Fig2}
			\end{figure}

			Therefore, we obtain the following solution for free energy local minima
			\begin{align}
				\begin{split}
					\frac{Q_\mathrm{NC}}{Q_\mathrm{C}}
					=f_{lmn}(Q_\mathrm{IC})
					\equiv\frac{Q_\mathrm{IC}+z_{lmn}}{Q_\mathrm{C}+z_{lmn}}.
				\end{split}
				\label{LocalMinimaModular}
			\end{align}
			Equation \eqref{LocalMinimaModular} is our central theoretical result. It means that if $Q_\mathrm{IC}$ and $Q_\mathrm{C}$ are given, then CDW phases between IC and C are described by distinct Eisenstein integers $z_{lmn}$.
			%For instance, $f_{100}$ corresponds to the NC phase in $1T$-TaS\textsubscript{2}. 
			%By fixing the two wavenumbers $Q_\mathrm{IC}$ and $Q_\mathrm{C}$, we obtain many other local minima (compare Figure \ref{FigMultivalley} (a), (d), (e), (f)).
			{Figure \ref{Fig2} shows the free energy local minima.} One can show that the set of transformations $f_{lmn}$ {have one-to-one correspondence with a} commutative subgroup of the projective linear group PSL$(2,\mathbb{Z}[\omega])$. {This group property implies that transition} between different local minima (i.e. different CDW phases) is obtained by a composition of $f_{lmn}$ and $f_{lmn}^{-1}$ (Figure \ref{Fig2} b,c). Three successive local minima lie on a circle whose center and radius can be calculated as in Supplementary Note \ref{SupplRadius}: These can explain successive CDW phase transition, such as the IC$\to$NC$\to$C phase transition in 1$T$-TaS\textsubscript{2} with decreasing temperature \cite{Multivalley}.
			{We also observe that the C wavenumbers $Q_\mathrm{C}$ are stationary points of the transformation, i.e. $f_{lmn}(Q_\mathrm{C})=1$ (Supplementary Video 1).}
			From this point of view, isotropic CDW phases can be understood clearly in a unified way. 
		This conformal method also applies to anisotropic stripe and T phases with minor modification (Supplementary Note \ref{SupplAnisotropic}).
	\section{Application to Experiments}
		We have shown that CDW phases can be described by a discrete conformal transformation. Which of the local minima become stable depends on many factors such as temperature and sample thickness, but our method is free from such detail. Therefore, our method can serve as the first step to analyze new CDW phases.
		
		\begin{figure}
			\begin{center}
				\includegraphics[width=0.65\linewidth]{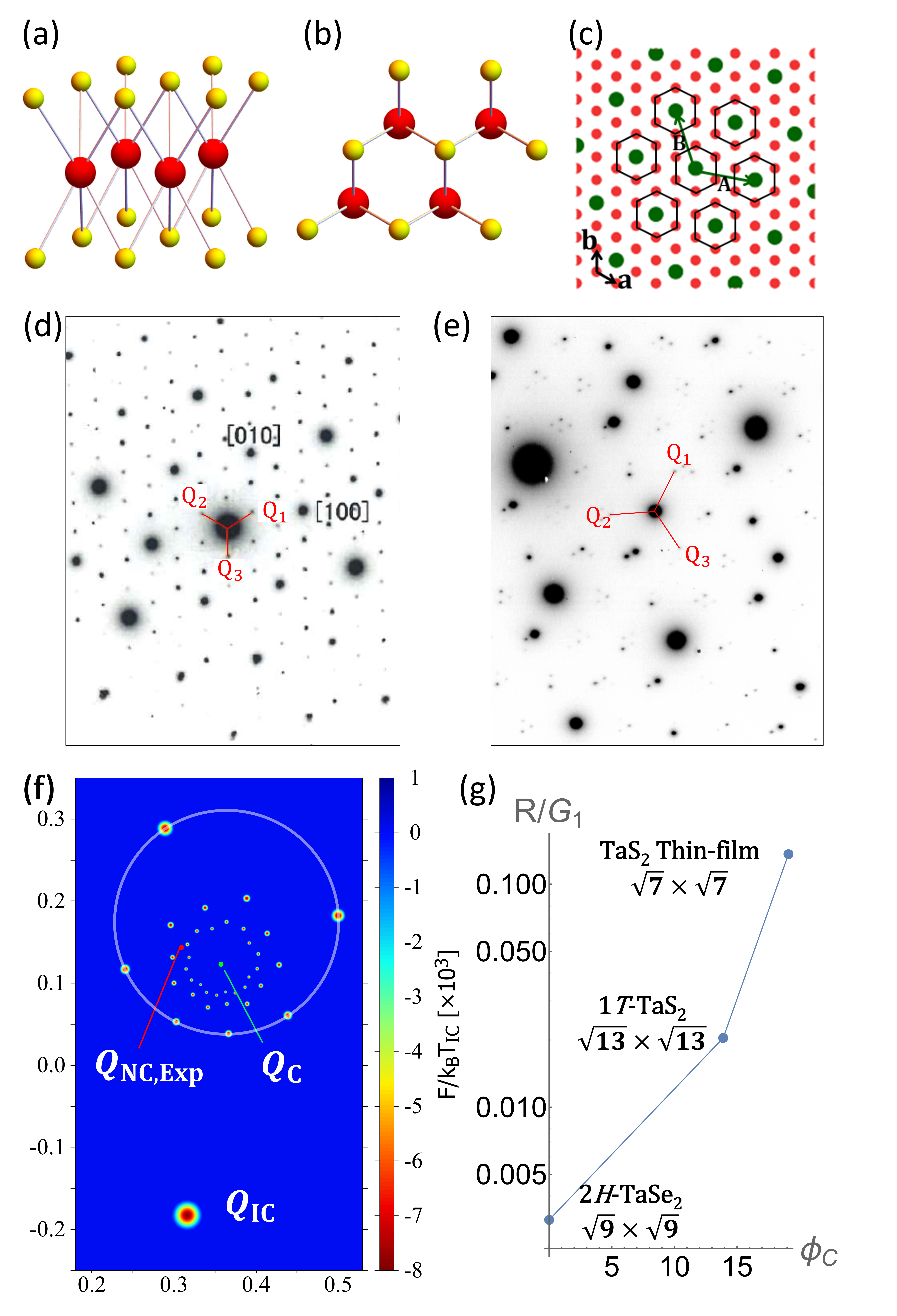}
			\end{center}
			\caption{(a) Structure of TaSe\textsubscript{2} thin-film. Red spheres represent Ta atoms and yellow spheres represent Se atoms. (b) Top-view of (a). (c) The $\sqrt{7}\times\sqrt{7}$ commensurate structure involving seven atoms, a central Ta atom (green) surrounded by six Ta atoms (red), in the unit cell. TED pattern of (d) $\sqrt{7}\times\sqrt{7}$ C phase \cite{Toshima} and (e) NC phase \cite{ToshimaThesis} in TaSe\textsubscript{2} thin-film. (f) The CDW free energy landscape for TaSe\textsubscript{2} thin film is calculated as a function of a general wavevector $\mathbf{Q}=(Q_x,Q_y)$ with $N=3$. See Supplementary Note \ref{SupplFlmn} for free energy parameters. The C phase is shown with a green dot. Experimental value of the NC phase is shown with a red dot. An enlarged image with local minima from Eq. \eqref{LocalMinimaModular} is shown in Supplementary Figure \ref{SuppFig3}. (g) Radius of circles shared by $N=1$ harmonics as a function of commensurate angle $\phi_\mathrm{C}$.}
			\label{TEM}
		\end{figure}
		We apply our analysis to a new experimental result. Ultra-thin sheets of TaSe\textsubscript{2} have been synthesized with ``de-chalcogenide method" \cite{Toshima}.
		The crystal structure is shown in Figure \ref{TEM} (a) and (b). Their transmission electron diffraction (TED) patterns (obtained at room temperature) are shown in Figure \ref{TEM} (d) and (e). Figure \ref{TEM} (d) is identified as the $\sqrt{7}\times\sqrt{7}$ {C phase} as shown in Figure \ref{TEM} (c) \cite{Toshima}.  In addition to the C phase, a new NC phase with $|\mathbf{Q}_\mathrm{NC}|/|\mathbf{G}_1|=0.341(2)$ and $\phi=25.0(2)^\circ$ was observed {in two different samples (Figure \ref{TEM} (e) \cite{ToshimaThesis} and Supplementary Figure 1)}. To explain the appearance of this experimental NC phase, we first use Eq. \eqref{LocalMinimaModular} to see which local minimum it corresponds to, then we calculate the CDW free energy using Eq. \eqref{FreeEnergy}.
		
		The C phase and the NC phase in TaS\textsubscript{2} thin films are obtained at room temperature, so $\mathbf{Q}_\mathrm{IC}$ with a much higher transition temperature could not be measured experimentally. However, we can obtain its approximate value based on the experimental NC phase. In $1T$-TaS\textsubscript{2} and $2H$-TaSe\textsubscript{2}, $\mathbf Q_\mathrm{C}$ and $\mathbf Q_\mathrm{IC}$ differ by no more than a few percent \cite{Wilson_Review_1975,NS1T,NS2H}: We assume that this is also true for our TaSe\textsubscript{2} thin-film.
		
		{Different MX\textsubscript{2} polytypes can have different nesting mechanisms. The IC phase in 2$H$-TaSe\textsubscript{2} can be formed by saddle point (van Hove singularity) nesting \cite{Rice-Scott,Neto}. In thinner systems, the attractive force caused by the van Hove singularity could be stronger than the force caused by Fermi surface nesting. Saddle point nesting can occur both in the $\Gamma$-M direction which is parallel to $G_1$, or in the $\Gamma$-K ($\Gamma$-K$'$) which is tilted $30^\circ$ ($-30^\circ$) from $G_1$. From Ref. \cite{Rice-Scott} one can estimate $|Q_\mathrm{IC}|=0.53$ in the $\Gamma$-M direction, but this value is too large: Experimentally, $|Q_\mathrm{C}|^{(i)}$ and $|Q_\mathrm{IC}|^{(i)}$ differ only by a few percent. On the other hand, in the $\Gamma$-K or $\Gamma$-K$'$ direction one can estimate $|Q_\mathrm{IC}|=0.31$ \cite{Rice-Scott,Neto}. } In fact, using Eq. \eqref{LocalMinimaModular}, we observe that the local minimum $f_{031}(Q_\mathrm{IC})$ corresponds to the experimental NC phase if we set $|Q_\mathrm{IC}|=0.35\sim0.38$ and $\phi=-30^\circ$ (Supplementary Video 2). The $-30^\circ$ angle implies that the IC phase in this sample is formed by van Hove singularity rather than Fermi-surface nesting.

		We also calculated the free energy \eqref{FreeEnergy} with temperature dependence of the form $a_0/2=T-T_\text{IC}$ where $T_\text{IC}$ is the IC-CDW transition temperature. The result of numerical calculation for TaSe\textsubscript{2} thin film is shown in Figure \ref{TEM}. Note the multivalley free-energy landscape which coincide with the local minima obtained by the conformal method. Therefore, our analysis can actually explain the appearance of the NC phase and agrees with experimental value as in previous studies \cite{NTYS,NS1T,NS2H,Multivalley}. 
		\\
		
	\section{Discussion}
		In this article we investigated the conformality of CDW in MX\textsubscript{2} compounds. {We showed that discommensurate CDW phases and successive IC-NC-C phase transition can be understood using discrete conformal transformations.}
		The discreteness of this transformation is due to commensurability of CDW with the underlying base lattice. We also applied our analysis to experimental results of a new NC phase in TaSe\textsubscript{2} thin-film with $\sqrt{7}\times\sqrt{7}$ commensurate structure. In what follows we give some implications of our results.

		First, note that $Q_\mathrm{C}$ is a stationary point of the transformation \eqref{LocalMinimaModular}.
		Therefore, one can regard $Q_\mathrm{IC}$ as a source, $Q_\mathrm{C}$ as a drain, and $Q_\mathrm{NC}$ as crossing points of the transformation.  From this point of view, discommensuration may be understood as deviation from fixed points. 
		This picture is a natural one because the $Q_\mathrm{IC}$ values {usually} determines $Q_\mathrm{C}$.
		
		Second, {in this article we described discommensurate CDW phases using the discrete conformal group PSL(2,$\mathbb{Z}[\omega]$), but a unified description of FQHL phases was also made using the discrete conformal group PSL(2,$\mathbb{Z}$) \cite{Belavin_Polyakov_Zamolodchikov,Laughlin,Kivelson,Conformal_Review}. These systems are both incompressible with a gapped spectrum. Therefore, we surmise that the field $\mathbb{K}$ in the group PSL(2,$\mathbb{K}$) is discrete for incompressible systems, and $\mathbb{K}$ is continuous for compressible systems. These are} summarized in Table \ref{SummaryTable}.
		
		Third, local minima with the same value of $|z_{lmn}|$ lie on a circle (Supplementary Note \ref{SupplRadius}). Figure 2(e), (f) and Figure 3 (f) show the case with $|z_{lmn}|=1$. Figure 3 (g) shows the radius $R$ of these circles as a function of the commensurate angle $\phi_\mathrm{C}$. Using experimental values of $Q$ we find that $R$ increase almost logarithmically with $\phi_C$. 
		The difference between $|Q_\mathrm{C}|$ and $|Q_\mathrm{IC}|$ in 1$T$-TaS\textsubscript{2} ($\sqrt{13}$ commensurate structure) and 2$H$-TaSe\textsubscript{2} ($\sqrt{9}$ commensurate structure) are both $2-3\%$, but the local minima are spread more widely (i.e. has larger $R$) in 1$T$-TaS\textsubscript{2}.
		On the other hand, NC is reported in 1$T$-TaS\textsubscript{2} but not in 2$H$-TaSe\textsubscript{2}. Moreover, our TaSe\textsubscript{2} thin film has large $R$ which allows $|Q_\mathrm{C}|$ and $|Q_\mathrm{IC}|$ to differ by more than $5\%$. Therefore, we surmise that, in addition to the norm of $Q$, the phase of $Q$ must be considered to obtain a stable NC phase.
		
		Finally, possible future studies are as follows. Further consequences of describing C-CDW wavevectors by Eisenstein integers will be discussed in a following article. We expect that our formalism is also applicable to other superlattice systems \cite{Bak_1982} and other triple-Q systems like Skyrmion lattices. Although we have focused on a triangular lattice, similar results may be obtained in square lattice \emph{mutatis-mutandis} using Gauss integers $\mathbb{Z}[i]$ with elements $z=a+\mathrm{i}b$, $a,b\in\mathbb{Z}$ to describe CDWs in CuO, FeTe, and charge order.

		\acknowledgments{
			{The authors thank the Supercomputer Center, the Institute for Solid State Physics, the University of Tokyo for the use of the facilities. T.N.I. was supported by JSPS KAKENHI Grant No. JP18K13495 and No. 21K13852.}
		}

		\begin{table}
			\caption{
				A comparison of 2D electron system and their representative groups. The incompressibility of a system is usually associated with the presence of a band gap. {We surmise that the field $\mathbb{K}$ in the group PSL(2,$\mathbb{K}$) is discrete for incompressible systems, and $\mathbb{K}$ is continuous for compressible systems.}
			}
			\label{SummaryTable}
			\begin{tabular}{|c|c|c|c|}
				\hline
				2D system&\tablesplit{Compressibility/}{gap type}&Material&Group
				\\
				\hline\hline
				IC\textsubscript{1}$\leftrightarrow$IC\textsubscript{2}&Compressible? \cite{Yang2000}&MX\textsubscript{2}&$\mathrm{PSL}(2,\mathbb{C})$
				\\
				\hline
				\tablesplit{IC$\leftrightarrow$NC}{NC\textsubscript{1}$\leftrightarrow$NC\textsubscript{2}}&\multirow{2}{*}{\begin{tabular}{c}\\[-1
				em]Incompressible \cite{Laughlin}\\ Comme. gap\end{tabular}}&\tablesplit{1T-TaS\textsubscript{2}}{TaSe\textsubscript{2} thin-film}&\multirow{2}{*}{\begin{tabular}{c}\\[0.35em]$\mathrm{PSL}(2,\mathbb{Z}[\omega])$\end{tabular}}
				\\ \cline{1-1} \cline{3-3}
				IC$\leftrightarrow$C&&\tablesplit{2H-TaSe\textsubscript{2}}{VSe\textsubscript{2}}&
				\\ \cline{1-1} \cline{3-3}
				\hline\hline
				QHL&\tablesplit{Incompressible}{Magneto-roton gap}&GaAs& PSL$(2,\mathbb{Z})$
				\\
				\hline
			\end{tabular}
		\end{table}
		\bibliographystyle{apsrev4-1}
		\bibliography{References}

	\renewcommand{\theequation}{S.\arabic{section}.\arabic{equation}}
	\renewcommand{\thesection}{\arabic{section}}
	\newpage
	\setcounter{section}{0}
	\refstepcounter{section}
	\setcounter{equation}{0}
	\section*{Supplementary Note \arabic{section}: Definition of reciprocal lattice vectors}
		\label{SupplG}
		{Consider a triangular lattice with primitive lattice vectors $\mathbf{a}_1=(1,0)a$ and $\mathbf{a}_2=\frac{1}{2}(-1,\sqrt{3})a$ which are separated by $120^\circ$. Reciprocal lattice vectors $\mathbf{b}_1$ and $\mathbf{b}_2$ which satisfy $\mathbf{a}_i\cdot\mathbf b_j=2\pi\delta_{ij}$ ($i,j=1,2$) are given by $\mathbf{b}_1=(1,1/\sqrt{3})/a$ and $\mathbf{b}_2=(0,2/\sqrt{3})/a$. $\mathbf{a}_i$ and $\mathbf{b}_i$ are not parallel but are separated by $30^\circ$. However, for the description of CDW we would like to have CDW wavevectors which are parallel to the CDW superlattice. Moreover, the triangular lattice and its reciprocal lattice both have three-fold rotation symmetry. So, it would be convenient to have a new set of reciprocal lattice vectors $\mathbf{G}_i$ which are parallel to $\mathbf{a}_i$. From the three-fold rotational symmetry of the system we require that $\mathbf{G}_1$ and $\mathbf{G}_2$ are separated by $120^\circ$ and we also define $\mathbf{a}_3=-\mathbf{a}_1-\mathbf{a}_2$, $\mathbf{G}_3=-\mathbf{G}_1-\mathbf{G}_2$. The two sets of reciprocal lattice vectors are related by $\mathbf{b}_1=2/3(\mathbf{G}_1-\mathbf{G}_3)$ and $\mathbf{b}_2=2/3(\mathbf{G}_2-\mathbf{G}_3)$. $\mathbf{G}_i$ ($i=1,2,3$) are normalized by the condition $\mathbf{a}_i\cdot\mathbf{G}_i=2\pi$. For $i\ne j$, $\mathbf{a}_i$ and $\mathbf{G}_j$ are mutually separated by $120^\circ$ which implies $\mathbf{a}_i\cdot\mathbf{G}_j=2\pi\cos(2\pi/3)=-\pi$. Consequently, we have the condition $\mathbf{a}_i\cdot\mathbf{G}_j=2\pi\delta_{ij}-\pi(1-\delta_{ij})=\pi(3\delta_{ij}-1)$.}
		
		\newpage
		\refstepcounter{section}
		\setcounter{equation}{0}
	\section*{Supplementary Note \arabic{section}: Circles shared by local minima}
		\label{SupplRadius}
		The M\"obius transformation is a conformal transformation which maps circles to circles. In particular, there exists a map between a circle containing local minima $(Q_1,Q_2,Q_3)$ to another circle containing different local minima $(Q_1',Q_2',Q_3')$. The center and radius of each circles can be calculated, for instance, by the transformation $F(Q)=\frac{Q-Q_1}{Q_2-Q_1}$
		which transforms the equation of a circle $|Q-z_0|=R^2$ to $|F(Q)-c|^2=r^2$ with $c=F(Q_0)$. Then, from $F(Q_1)=0$, $F(Q_2)=1$, and $F(Q_3)=\frac{Q_3-Q_1}{Q_2-Q_1}\equiv\chi$, one can show that $c=\frac{\chi-|\chi|^2}{\chi-\chi^\ast}$, $Q_0=F^{-1}(c)=(Q_2-Q_1)c+Q_1$, and $R^2=|(Q_2-Q_1)c|^2$. For the circle containing the first harmonics ($N=1$) we obtain
		\begin{equation}
			{Q_0=\frac{Q_\mathrm{C}-Q_\mathrm{IC}|Q_\mathrm{C}|^2}{1-|Q_\mathrm{C}|^2},\quad R^2=\frac{|Q_\mathrm{C}-Q_\mathrm{IC}|^2|Q_\mathrm{C}|^2}{(1-|Q_\mathrm{C}|^{2})^2}.}
			\label{Eqn_Circle}
		\end{equation}

		\newpage
		\refstepcounter{section}
		\setcounter{equation}{0}
	\section*{Supplementary Note \arabic{section}: Domain walls of anisotropic CDW}
		\label{SupplAnisotropic}
		Anisotropic CDW phases have domain walls which are tilted from each other by an angle that is different from $120^\circ$. {This angle can be obtained with good accuracy using real-spacing imaging such as scanning tunnelling microscope, so we use it as input parameter.}
		Quasi-stripe (T) and stripe phases need to be treated separately.

		\subsection{Quasi-stripe phases}
		Consider CDW harmonic wavevectors $k^{(1)}, k^{(2)}, k^{(3)}$ which are related by $k^{(1)}=w_1k^{(1)}$, $k^{(2)}=w_2k^{(1)}$, and $k^{(3)}=w_3k^{(1)}$. Here, $w_1=1$ is introduced for convenience such that our definition of $w_i$ gives
		\begin{equation}
			\frac{k^{(1)}}{w_1}=\frac{k^{(2)}}{w_2}=\frac{k^{(3)}}{w_3}.
		\end{equation}
		For isotropic phases we have $w_2=\omega$ and $w_3=\omega^2$. For anisotropic phases, we can write $k^{(j)}/k^{(i)}=e^{\mathrm{i}\phi_{ij}}|k^{(j)}|/|k^{(i)}|$ and observe that the oriented angle $\phi_{ij}$ between $k^{(i)}$ and $k^{(j)}$ and their norm are related by the law of sines
		\begin{equation}
			\frac{|k^{(1)}|}{\sin\phi_{23}}=\frac{|k^{(2)}|}{\sin\phi_{31}}=\frac{|k^{(3)}|}{\sin\phi_{13}}
		\end{equation}
		Consequently, we obtain $w_2=e^{\mathrm{i}\phi_{12}}\frac{\sin\phi_{13}}{\sin\phi_{32}}$, $w_3=e^{\mathrm{i}\phi_{13}}\frac{\sin\phi_{12}}{\sin\phi_{23}}$.

		Next, the commensurability condition $k^{(i)}=\mu q^{(i)}-\nu q^{(i+1)}$ between domain wall wavevectors $q^{(i)}$ and CDW harmonics $k^{(i)}$ is equivalent to the commensurate structure
		\begin{equation}
			q^{(i)}=\frac{\mu^2k^{(i)}+\mu\nu k^{(i+1)}+\nu^2k^{(i+2)}}{\mu^3-\nu^3}=\frac{\mu^2w_i+\mu\nu w_{i+1}+\nu^2w_{i+2}}{\mu^3-\nu^3}k^{(1)}.\label{qEqn}
		\end{equation}
		This equation is equivalent to
		\begin{align}
			\frac{k^{(i)}}{w_i}&=\frac{\mu^3-\nu^3}{\mu^2w_i+\mu\nu w_{i+1}+\nu^2 w_{i+2}}q^{(i)}.
		\end{align}

		Now, $Q_{lmn}^{(i)}$ in the Nakanishi-Shiba expansion $\psi_i(\mathbf r)=\sum_{lmn}\Delta_{lmn}^{(i)}\exp\{\mathrm{i} \mathbf Q_{lmn}^{(i)}\cdot\mathbf r\}$ can be written as
		\begin{align}
			Q_{lmn}^{(i)}&=Q^{(i)}+lk^{(i)}+mk^{(i+1)}+nk^{(i+2)}
			\\
			&=Q^{(i)}+(lw_i+mw_{i+1}+nw_{i+2})\frac{k^{(i)}}{w_i}
			\\
			&=Q^{(i)}+z_{i,lmn}\omega^{i-1}(Q^{(i)}/Q_\mathrm{C}^{(i)}-1).
			\\
			&=\frac{Q^{(i)}}{Q_\mathrm{C}^{(i)}}(Q_\mathrm{C}^{(i)}+z_{i,lmn}\omega^{i-1})-z_{i,lmn}\omega^{i-1}.
		\end{align}
		In the third line we defined
		\begin{equation}
			z_{i,lmn}=\frac{lw_i+mw_{i+1}+nw_{i+2}}{\mu^2w_i+\mu\nu w_{i+1}+\nu^2 w_{i+2}}\frac{\mu^3-\nu^3}{\mu-\nu\omega}
		\end{equation}
		and in the last line we rearranged the equation using $Q_\mathrm{C}^{(i)}=\omega^{(i-1)}/(\mu-\nu\omega)$.
		For isotropic phases we obtain $z_{i,lmn}=l+m\omega+n\omega^2$.
		Consequently, the $Q$-dependent part of the free energy can be written as 
		\begin{align}
			f(Q)&\equiv\sum_{i=1}^3\sum_{lmn}\left|\Delta_{lmn}^{(i)}\right|^2\left|Q_{lmn}^{(i)}-Q_\mathrm{IC}^{(i)}\right|^2
			\\
			&=\sum_{i=1}^3\sum_{lmn}\left|\Delta_{lmn}^{(i)}\right|^2\left|\frac{Q^{(i)}}{Q_\mathrm{C}^{(i)}}(Q_\mathrm{C}^{(i)}+z_{i,lmn}\omega^{i-1})-(Q_\mathrm{IC}^{(i)}+z_{i,lmn}\omega^{i-1})\right|^2.
		\end{align}
		Regarding $\Delta^{(i)}_{lmn}$ as different degrees of freedom, the free energy is minimized if components with different values of $l,m,n$ vanish independently, that is if
		\begin{equation}
			\frac{Q^{(i)}}{Q_\mathrm{C}^{(i)}}=\frac{Q_\mathrm{IC}^{(i)}+z_{i,lmn}\omega^{i-1}}{Q_\mathrm{C}^{(i)}+z_{i,lmn}\omega^{i-1}}.
		\end{equation}

		\subsection{Stripe phases}
		Consider the stripe CDW phase with domain wall wavevectors $q^{(1)}=0$, $q^{(2)}=-\mathrm{i}|q|e^{\mathrm{i}\theta}$, and $q^{(3)}=\mathrm{i}|q|e^{\mathrm{i}\theta}$. In this case, Eq. \eqref{qEqn} cannot be used and we need to treat stripe phases separately. Substitute $q^{(i)}$ to the Q-dependent part of the free energy and we obtain
		\begin{align*}
			f(Q)&\equiv\sum_{i=1}^3\sum_{lmn}\left|\Delta_{lmn}^{(i)}\right|^2\left|Q_{lmn}^{(i)}-Q_\mathrm{IC}^{(i)}\right|^2
			\\
			&=\sum_{lmn}\left|
				\Delta_{lmn}^{(1)}
			\right|^2\left|
			(n-m)\mathrm{i}|q|e^{\mathrm{i}\theta}+Q_\mathrm{C}^{(1)}-Q_\mathrm{IC}^{(1)}
			\right|^2
			\\
			&+\sum_{lmn}\left|
				\Delta_{lmn}^{(2)}
			\right|^2\left|
			(m-l-1)\mathrm{i}|q|e^{\mathrm{i}\theta}+Q_\mathrm{C}^{(2)}-Q_\mathrm{IC}^{(2)}
			\right|^2
			\\
			&+\sum_{lmn}\left|
				\Delta_{lmn}^{(3)}
			\right|^2\left|
			-(n-l-1)\mathrm{i}|q|e^{\mathrm{i}\theta}+Q_\mathrm{C}^{(3)}-Q_\mathrm{IC}^{(3)}
			\right|^2.
		\end{align*}
		Regarding $\Delta^{(i)}_{lmn}$ as different degrees of freedom, the free energy is minimized if components with different values of $l,m,n$ vanish independently, that is if
		\begin{align}
			(n-m)\left\{
				|q|(n-m)+\mathrm{Im}\left[e^{-\mathrm{i}\theta}\left(Q_\mathrm{C}^{(1)}-Q_\mathrm{IC}^{(1)}\right)\right]
			\right\}=0,\label{cond1}
			\\
			(m-l-1)\left\{
				|q|(m-l-1)+\mathrm{Im}\left[\omega e^{-\mathrm{i}\theta}\left(Q_\mathrm{C}^{(1)}-Q_\mathrm{IC}^{(1)}\right)\right]
			\right\}=0,\label{cond2}
			\\
			(n-l-1)\left\{
				|q|(n-l-1)-\mathrm{Im}\left[\omega^2 e^{-\mathrm{i}\theta}\left(Q_\mathrm{C}^{(1)}-Q_\mathrm{IC}^{(1)}\right)\right]
			\right\}=0.\label{cond3}
		\end{align}
		Stripe phases satisfy this set of equations.
		We solve it for the case $n-m=0$. If $m-l-1=0$ then $|q|$ is arbitrary. On the other hand, if $m-l-1\ne0$, we obtain
		\begin{align}
			(m-l-1)\left\{
				|q|(m-l-1)+\mathrm{Im}\left[\omega e^{-\mathrm{i}\theta}\left(Q_\mathrm{C}^{(1)}-Q_\mathrm{IC}^{(1)}\right)\right]
			\right\}=0,
			\\
			(m-l-1)\left\{
				|q|(m-l-1)-\mathrm{Im}\left[\omega^2 e^{-\mathrm{i}\theta}\left(Q_\mathrm{C}^{(1)}-Q_\mathrm{IC}^{(1)}\right)\right]
			\right\}=0.
		\end{align}
		Subtracting these equations, we obtain
		\begin{equation}
			\mathrm{Im}\left[e^{-\mathrm{i}\theta}\left(Q_\mathrm{C}^{(1)}-Q_\mathrm{IC}^{(1)}\right)\right]=0\Rightarrow e^{-\mathrm{i}\theta}\left(Q_\mathrm{C}^{(1)}-Q_\mathrm{IC}^{(1)}\right)=\pm\left|Q_\mathrm{C}^{(1)}-Q_\mathrm{IC}^{(1)}\right|,
		\end{equation}
		where the sign is chosen such that $|q|$ is positive. This result implies that stripe wave vectors are given by
		\begin{equation}
			|q|=\frac{\sqrt{3}}{2}\left|\frac{Q_\mathrm{C}^{(1)}-Q_\mathrm{IC}^{(1)}}{m-l-1}\right|,
		\end{equation}
		where we have used $\mathrm{Im}[\omega]=\sqrt{3}/2$.
		\newpage
		\refstepcounter{section}
		\setcounter{equation}{0}
	\section*{Supplementary Note \arabic{section}: Type 1 and Type 2 Free Energies}
		\label{SupplType}
		We can use two types of free energies with different forms of $e_i(\mathbf Q^{(i)})$. The first one (type 1) is the free energy considered by Nakanishi and Shiba \cite{NS1T} which uses phenomenological parameters which reproduce a ring-like diffuse scattering obtained in  and electron diffraction experiment:
		\begin{align}
			e_i(\mathbf Q^{(i)})=(1-\xi_i)u+v\xi_i(1-\cos(6\phi_i))
			\qquad \mbox{(Type 1)}
			\nonumber
		\end{align}
		where
		\begin{align*}
			\xi_i=\left\{
			\begin{array}{ll}
				1-\frac{s(|\mathbf{Q}^{(i)}|-|\mathbf Q_\text{IC}^{(i)}|)^2}{|\mathbf G_i|^2},&\mbox{ if }\frac{(|\mathbf{Q}^{(i)}|-|\mathbf Q_\text{IC}^{(i)}|)^2}{|\mathbf G_i|^2}<s^{-1}
				\\
				0,&\mbox{ otherwise},
			\end{array}
			\right.
		\end{align*}
		and $u$ and $v$ are parameters; $\phi_i$ is the angle between $\mathbf Q^{(i)}$ and $\mathbf G_i$, $\mathbf Q_\text{IC}^{(i)}$ is the fundamental wavevector of the IC phase. This $e_i(\mathbf Q^{(i)})$ has a circular ditch with the radius $\mathbf Q_\text{IC}^{(i)}$ around $\mathbf Q^{(i)}=0$.
		The second one (type 2) uses McMillan's free energy \cite{McMillan3} which has the gradient term
		\begin{align*}
			\int& d^2r\psi_i^\ast e_i(-\mathrm{i}\nabla)\psi_i
			\\
			&=
			\int d^2r\left[e|(\mathbf Q_\text{IC}^{(i)}\cdot\nabla-\mathrm{i}|\mathbf Q_\text{IC}^{(i)}|^2)\psi_i|^2+f|\mathbf Q_\text{IC}^{(i)}\times\nabla\psi_i|^2\right]
			\\
			&=\int d^2r\psi_i^\ast\left[e(\mathbf Q_\text{IC}^{(i)}\cdot\nabla-\mathrm{i}|\mathbf Q_\text{IC}^{(i)}|^2)^2+f|\mathbf Q_\mathrm{IC}^{(i)}\times\nabla|^2\right]\psi_i
			\\
			&=\int d^2r\psi_i^\ast\left[e|\mathbf Q_\text{IC}^{(i)}|^2|\nabla-\mathrm{i}\mathbf Q_\text{IC}^{(i)}|^2+(f-e)|\mathbf Q_\text{IC}^{(i)}\times\nabla|^2\right]\psi_i,
		\end{align*} 
		therefore
		\begin{equation}
			e_i(-\mathrm{i}\nabla)=e|\mathbf Q_\text{IC}^{(i)}|^2|(-\mathrm{i}\nabla)-\mathbf Q_\text{IC}^{(i)}|^2+(f-e)|\mathbf Q_\text{IC}^{(i)}\times(-\mathrm{i}\nabla)|^2
			\label{eiDef}
		\end{equation}
		where $e$ and $f$ are constants, the first equality is McMillan's original definition, the second equality follows from integration by parts assuming that $\psi_i$ are non-decaying periodically oscillating functions, and in the third equality we did some rearrangement using trigonometric identities.

		Now, assuming $e=f=s/(|\mathbf Q_\text{IC}^{(i)}|^2|\mathbf G_i|^2)$ we obtain the type 2 free energy with
		\begin{align}
			e_i(\mathbf{Q})=s|\mathbf{Q}-\mathbf Q_\text{IC}^{(i)}|^2/|\mathbf G_i|^2.\qquad\mbox{(Type 2)}
			\nonumber
		\end{align}
		This $e_i(\mathbf{Q})$ has a circular (or elliptical if $e\ne f$) bulge around $\mathbf Q_\text{IC}^{(i)}$.
		In their study of 2\textit{H}-TaSe\textsubscript{2}, Nakanishi and Shiba obtained this form as an approximate form of type 1 free energy with $\mathbf{Q}^{(i)}\approx\mathbf Q_\text{IC}^{(i)}$ \cite{NS2H}, but it is actually also a special case of McMillan's original definition of free energy. Therefore, assuming $e=f$, the type 2 free energy can be used for general values of $\mathbf Q^{(i)}$.

		\newpage
		\refstepcounter{section}
		\setcounter{equation}{0}
	\section*{Supplementary Note \arabic{section}: Calculation of Free Energy}
		\label{SupplFlmn}
		The free energy \eqref{F2} can be integrated using Eq. \eqref{philmn}. First, we expand the $\alpha^2$ term in Eq. \eqref{FreeEnergy}. Using $a(\mathbf r)=a_0+2a_1\text{Re}\sum_{i=1}^3e^{\mathrm{i}\mathbf G_i\cdot\mathbf r}$, $\alpha(\mathbf r)=\text{Re}\sum_{i=1}^3\psi_i(\mathbf r)$, and rewrite Eq. \eqref{philmn} into the form $\psi_i(\mathbf r)=e^{\mathrm{i} \mathbf Q_\text{C}^{(i)}\cdot\mathbf r}\phi_i(\mathbf r)$, we obtain
		\begin{align*}
			\int d^2r\ a(\mathbf r)\alpha^2(\mathbf r)=a_0\int d^2r\left[\text{Re}\sum_{i=1}^3e^{\mathrm{i}\mathbf Q_\text{C}^{(i)}\cdot\mathbf r}\phi_i(\mathbf r)\right]^2
			+a_1\int d^2r\sum_{i=1}^3e^{i\mathbf G_i\cdot\mathbf r}\left[\text{Re}\sum_{j=1}^3e^{\mathrm{i}\mathbf Q_\text{C}^{(j)}\cdot\mathbf r}\phi_j(\mathbf r)\right]^2.
		\end{align*}
		Oscillating terms vanish after integration, so the non-zero terms are those with absolute values $|\phi_i|^2$:
		\begin{align*}
			\int d^2r\ a(\mathbf r)\alpha^2(\mathbf r)=\frac{a_0}{2}\sum_{i=1}^3\int d^2r\ |\phi_i|^2.
		\end{align*}
		Similarly, the $\alpha^3$ term is
		\begin{align*}
			\int d^2r\ b(\mathbf r)\alpha^3(\mathbf r)=-\int &d^2r\bigg\{\frac{3b_0}{4}(\phi_1\phi_2\phi_3+\text{c.c.})
			+\frac{3b_1}{8}\sum_{i=1}^3(\phi_i^2\phi_{i+1}^\ast+\text{c.c.})\bigg\}
		\end{align*}
		and the $\alpha^4$ term is
		\begin{align*}
			\int d^2r\ c(\mathbf r)\alpha^4(\mathbf r)
			=\int d^2r\sum_{i=1}^3\bigg\{\frac{c_0}{8}(3|\phi_i|^4+12|\phi_i\phi_{i+1}|^2)
			+\frac{3c_1}{4}(\phi_i\phi_{i+1}^{\ast2}\phi_{i+2}^\ast+\text{c.c.})
			+\frac{c_1}{4}(\phi_i^3\phi_{i+2}+\text{c.c.})\bigg\}.
		\end{align*}
		The term proportional to $b_0$ remains from the triple-$\mathbf{Q}$ condition $\mathbf Q_\text{C}^{(1)}+\mathbf Q_\text{C}^{(2)}+\mathbf Q_\text{C}^{(3)}=\mathbf 0$. The terms proportional to $a_1$, $b_1$ and $c_1$ are the Umklapp terms that remain from the commensurate structure $2 \mathbf Q_\text{C}^{(i)}-\mathbf Q_\text{C}^{(i+1)}=\mathbf G_i$.
		After some rearrangement we obtain
		\begin{align}
			F=&\int d^2r(f_\mathrm{Elastic}+f_\mathrm{Triple-Q}+f_\mathrm{Comme.}),\label{F2}
			\\
			f_\mathrm{Elastic}=&\sum_{i=1}^3[\phi_i^\ast A_i(\mathbf Q_\text{C}^{(i)}-i\nabla)\phi_i+B|\phi_i|^4+C|\phi_i\phi_{i+1}|^2],\nonumber
			\\
			f_\mathrm{Triple-Q}&=\frac{D}{2}(\phi_1\phi_2\phi_3+\text{c.c.}),\nonumber
			\\
			f_\mathrm{Comme.}=&\frac{1}{2}\sum_{i=1}^3[U\phi_i^2\phi_{i+1}^\ast+3Z\phi_i\phi_{i+1}^{\ast 2}\phi_{i+2}^\ast
			+Z\phi_i^3\phi_{i+2}+\text{c.c.}],
			\nonumber
		\end{align}
		where $A_i(\mathbf Q)=\frac{1}{2}a_0+e_i(\mathbf Q)$, $B=\frac{3}{8}c_0$, $C=\frac{3}{2}c_0+d_0$, $D=-\frac{3}{2}b_0$, $U=-\frac{3}{4}b_1$, $Z=\frac{1}{2}c_1$.

		Substitute Eq. \eqref{philmn} into Eq. \eqref{F2} and we obtain
		\begin{align*}
			F_\mathrm{Elatsic}&=\sum_{i=1}^3\sum
			\Delta_{lmn}^{(i)\ast}\Delta_{lmn}^{(i)}A_i(\mathbf{Q}_{lmn}^{(i)})
			\\
			&+B\sum_{i=1}^3\sum
			\Delta_{lmn}^{(i)\ast}\Delta_{l'm'n'}^{(i)}\Delta_{l''m''n''}^{(i)\ast}\Delta_{l'''m'''n'''}^{(i)}\delta_{l-l'+l''-l''',m-m'+m''-m''',n-n'+n''-n'''}
			\\
			&+C\sum_{i=1}^3\sum
			\Delta_{lmn}^{(i)\ast}\Delta_{l'm'n'}^{(i)}\Delta_{l''m''n''}^{(i+1)\ast}\Delta_{l'''m'''n'''}^{(i+1)}\delta_{l-l'+n''-n''',m-m'+l''-l''',n-n'+m''-m'''},
			\\\\
			F_\mathrm{Triple-Q}&=D\sum
			\Delta_{lmn}^{(1)}\Delta_{l'm'n'}^{(2)}\Delta_{l''m''n''}^{(3)}\delta_{l+n'+m'',m+l'+n'',n+m'+l''},
			\\\\
			F_\mathrm{Comme.}&=U\sum_{i=1}^3\sum
			\Delta_{lmn}^{(i)}\Delta_{l'm'n'}^{(i)}\Delta_{l''m''n''}^{(i+1)\ast}\delta_{l+l'-n''+1,m+m'-l'',n+n'-m''}
			\\
			&+3Z\sum_{i=1}^3\sum\Delta_{lmn}^{(i)}\Delta_{l'm'n'}^{(i+1)\ast}\Delta_{l''m''n''}^{(i+1)\ast}\Delta_{l'''m'''n'''}^{(i+2)\ast}\delta_{l-n'-n''-m'''+1,m-l'-l''-n''',n-m'-m''-l'''}
			\\
			&+Z\sum_{i=1}^3\sum\Delta_{lmn}^{(i)}\Delta_{l'm'n'}^{(i)}\Delta_{l''m''n''}^{(i)}\Delta_{l'''m'''n'''}^{(i+2)}\delta_{l+l'+l''+m'''+1,m+m'+m''+n''',n+n'+n''+l'''},
		\end{align*}
		where $\delta_{a,b,c}$ is the generalized delta function with $\delta_{a,b,c}=1$ if $a=b=c$ and $\delta_{a,b,c}=0$ otherwise. The sum $\sum$ is taken over all relevant positive integers $l,m,n,l',m',n',\dots$ with the condition $l\cdot m\cdot n=l'\cdot m'\cdot n'=l''\cdot m''\cdot n''=l'''\cdot m'''\cdot n'''=0$. Numerically, the free energy is minimized by $\partial F/\partial\Delta_{lmn}^{(i)}=0$ and setting a cutoff $l,m,n\leq N$. We choose the parameters $B=\frac{3}{8}c_0=1$, $C=\frac{3}{2}c_0+d_0=2$, $D=-\frac{3}{2}b_0=-0.1$, $U=-\frac{3}{4}b_1=-0.2$, $Z=\frac{1}{2}c_1=-0.5$, and $s=500$

		\newpage
		\refstepcounter{section}
		\setcounter{equation}{0}
	\section*{Supplementary Note \arabic{section}: Solution of Eq. (9)}
		\label{SupplMinEqn}
		{
			We solve Eq. (9) which is given by
			\begin{align*}
				\frac{\partial^2 F}{\partial\Delta^{(i)}_{lmn}\partial \mathbf{Q}^{(i)}}
				\propto\Delta_{lmn}^{(i)\ast}
				\frac{\partial e_i(\mathbf{Q}_{lmn}^{(i)})}{\partial\mathbf{Q}^{(i)}}=\mathbf{0}.
			\end{align*}
			where
			\begin{align*}
				e_i(\mathbf{Q})&=\frac{s|\mathbf{Q}-\mathbf Q_\text{IC}^{(i)}|^2}{|\mathbf G_i|^2}
				\\
				\mathbf Q_{lmn}^{(i)}&=\mathbf Q^{(i)}+l\mathbf k^{(i)}+m\mathbf k^{(i+1)}+n\mathbf k^{(i+2)}
				\\
				\mathbf k^{(i)}&=(\mu-\nu\omega)(\mathbf{Q}^{(i)}-\mathbf{Q}^{(i)}_\mathrm{C}).
			\end{align*}
			Equation (8) can be solved directly component-wise. However, if we assume three-fold rotational symmetry, then an easier method to solve Eq. (8) is to represent wavevectors by complex numbers. From the correspondence $\mathbf{Q}=(Q_x,Q_y)\to Q=Q_x+\mathrm{i}Q_y$ given in the main text we obtain
			\begin{align*}
				\frac{\partial e_i(\mathbf{Q}_{lmn}^{(i)})}{\partial\mathbf{Q}^{(i)}}&=\frac{\partial e_i(Q_{lmn}^{(i)})}{\partial Q^{(i)^\ast}}=0
				\\
				e_i(\mathbf{Q})&=e_i(Q)=s(Q-Q_\text{IC}^{(i)})(Q^\ast-Q_\text{IC}^{(i)\ast})/|G_i|^2
				\\
				Q_{lmn}^{(i)}&=Q^{(i)}+(l+m\omega+n\omega^2)k^{(i)}
				\\
				k^{(i)}&=(\mu-\nu\omega)(Q^{(i)}-Q^{(i)}_\mathrm{C}).
			\end{align*}
			where $\omega=e^{2\pi \mathrm{i}/3}$. The last two equations and $Q_\mathrm{C}^{(i)}=\omega^{i-1}|G_i|/(\mu-\nu\omega)$ imply
			\begin{equation}
				Q_{lmn}^{(i)}=[1+(l+m\omega+n\omega^2)(\mu-\nu\omega)]Q^{(i)}-(l+m\omega+n\omega^2)\omega^{i-1}|G_i|.
			\end{equation}
			Then, holomorphicity $\partial Q/\partial Q^\ast=0$ implies
			\begin{align*}
				\frac{\partial e_i(Q_{lmn}^{(i)})}{\partial Q^{(i)^\ast}}&=\frac{s(Q_{lmn}^{(i)}-Q_\text{IC}^{(i)})}{|G_i|^2}\frac{\partial(Q_{lmn}^{(i)\ast}-Q_\text{IC}^{(i)\ast})}{\partial Q^{(i)^\ast}}
				\\
				&=\frac{s(Q_{lmn}^{(i)}-Q_\text{IC}^{(i)})}{|G_i|^2}[1+(l+m\omega+n\omega^2)(\mu-\nu\omega)]^\ast
				\\
				&=0
			\end{align*}
			$l,m,n$ are positive integers by definition and we are assuming $\mu>1$. Therefore, $[1+(l+m\omega+n\omega^2)(\mu-\nu\omega)]\ne0$ which implies $Q_{lmn}^{(i)}-Q_\text{IC}^{(i)}=0$.
		}

		\renewcommand{\figurename}{Supplementary Figure}
		\setcounter{figure}{0}
		\newpage
	\begin{figure}[h]
		\includegraphics[width=\linewidth]{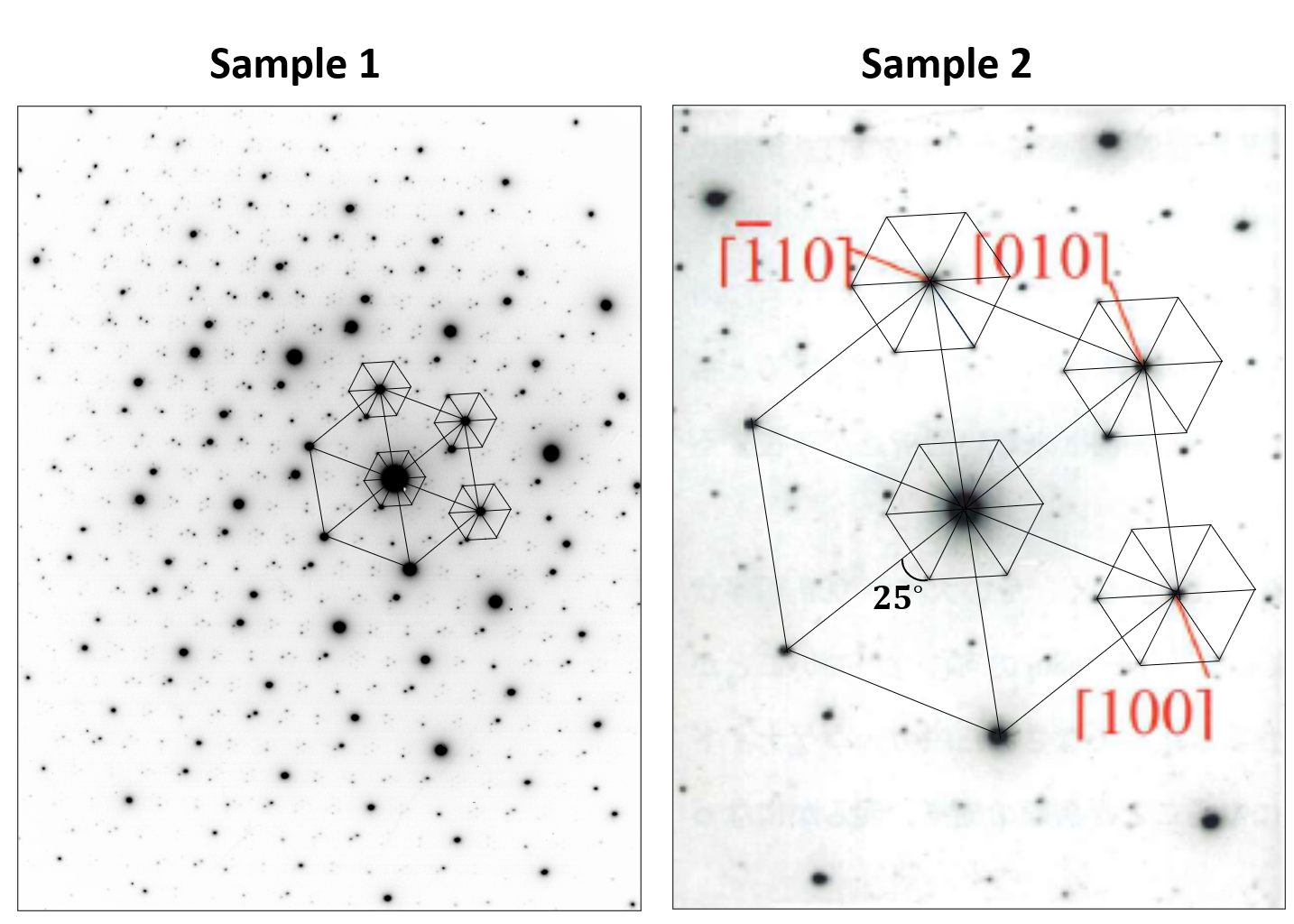}
		\caption{{The NC phase with $|\mathbf{Q}_\mathrm{NC}^{(i)}|/|\mathbf{G}_i|=0.341(2)$ and $\phi=25.0(2)^\circ$ was observed in two different samples.}}
		\label{SuppFig1}
	\end{figure}

	\newpage
	\begin{figure}[h]
		\includegraphics[width=\linewidth]{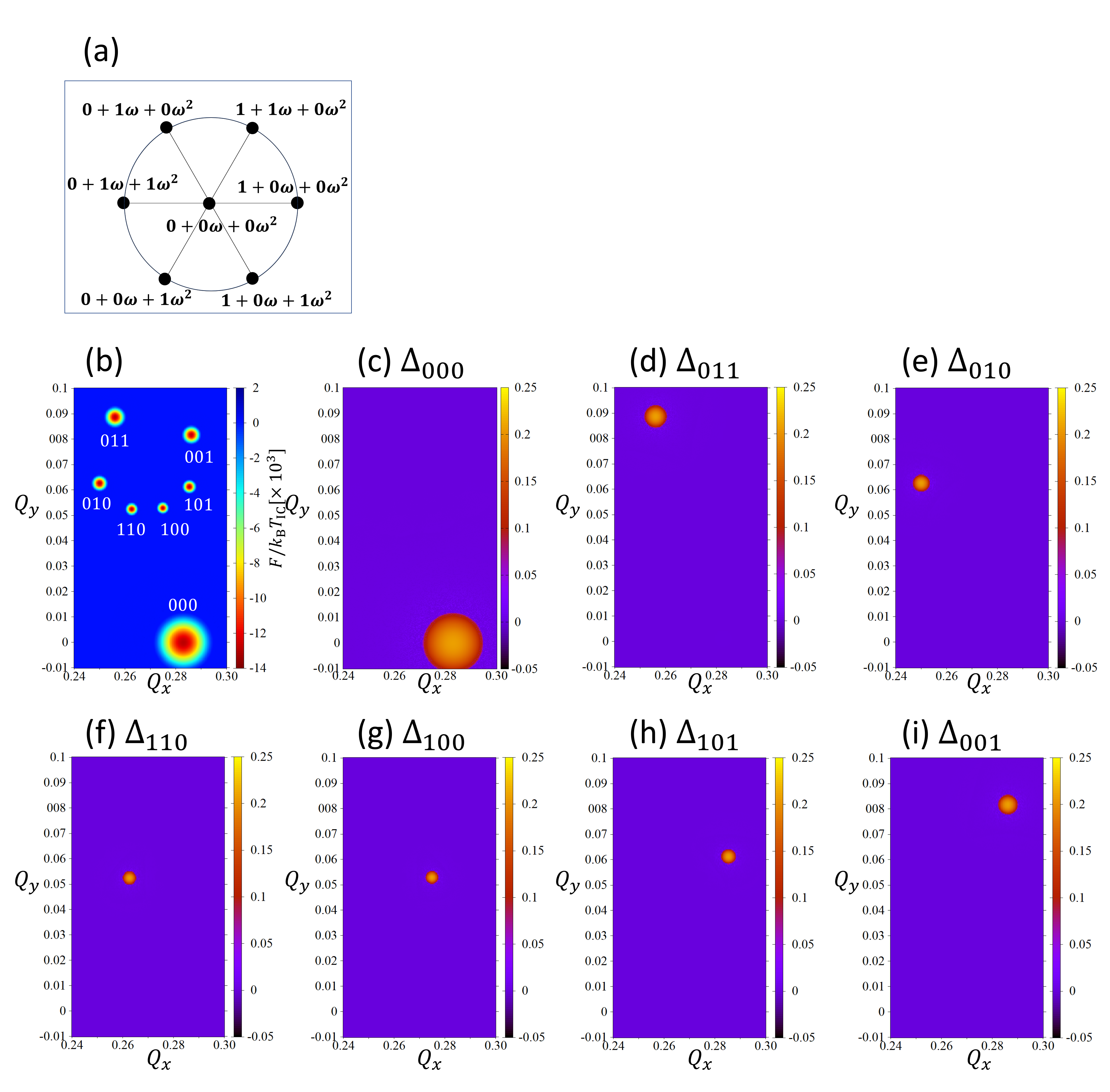}
		\caption{
			{(a) Eisenstein integers $z_{lmn}=l+m\omega+n\omega^2$ with $|z_{lmn}|\leq1$. (b) CDW free energy of 1$T$-TaS\textsubscript{2} as a function of order parameter $\psi_i(\mathbf r)=\sum_{l,m,n\geq0, l\cdot m\cdot n=0}\Delta_{lmn}^{(i)}\exp\{\mathrm{i} \mathbf Q_{lmn}^{(i)}\cdot\mathbf r\}$ where different $l,m,n$ label different harmonics. The sum $\sum_{l,m,n\geq0, l\cdot m\cdot n=0}$ is actually a sum over all Eisenstein integers $\sum_{z_{lmn}\in\mathbb{Z}[\omega]}$, so each harmonics correspond to different Eisenstein integers. (c)-(i) show that $\Delta_{lmn}$ has largest amplitude at the position of local minima in (b). Therefore, we conclude that each CDW local minima corresponds to a different value of $z_{lmn}$ with a good accuracy.
			}
		}
		\label{SuppFig2}
	\end{figure}

	\begin{figure}[h]
		\includegraphics[width=\linewidth]{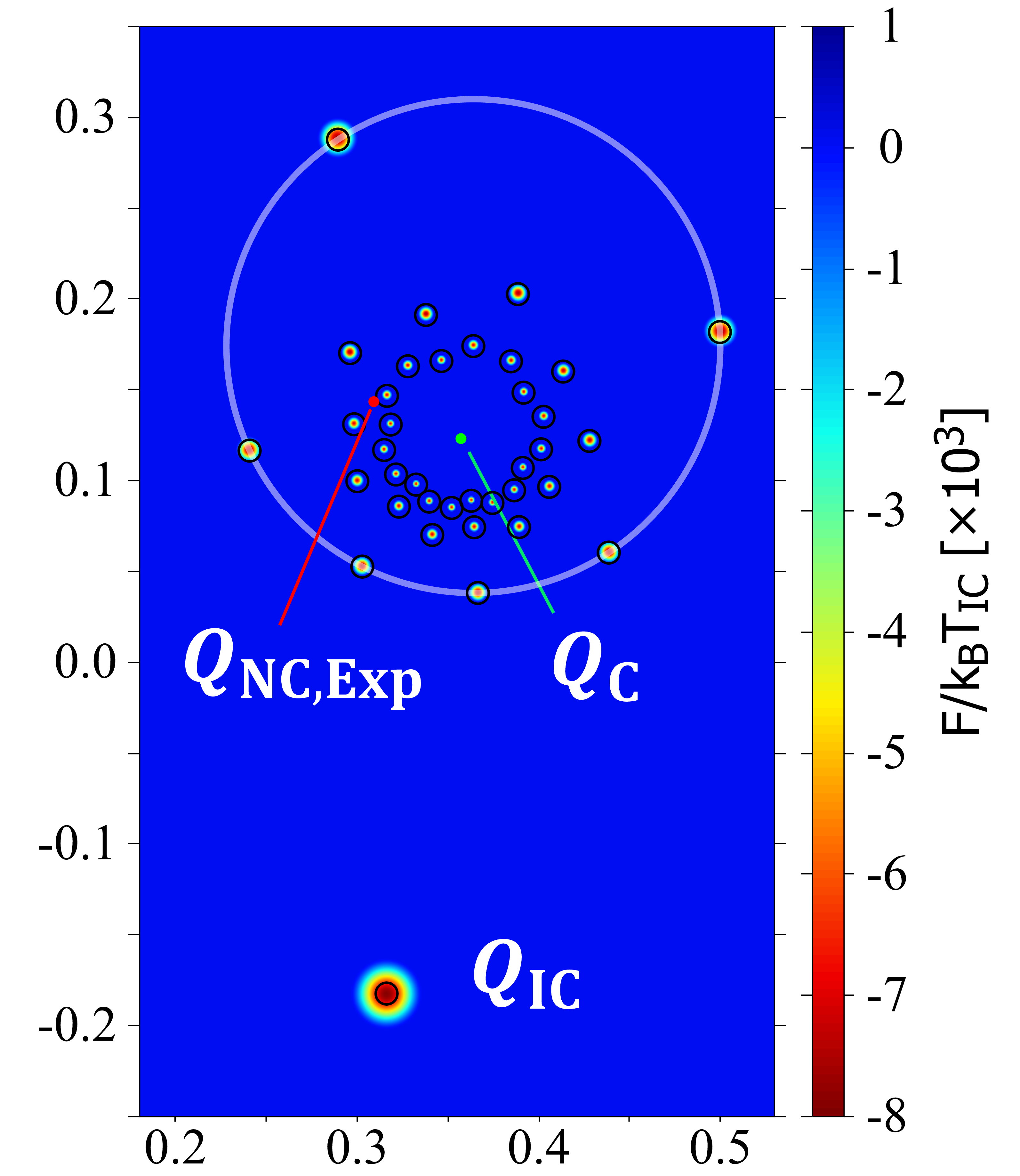}
		\caption{
			{Enlarged image of CDW free energy landscape for TaSe\textsubscript{2} thin film. The black circles show CDW local minima obtained from Eq. \eqref{LocalMinimaModular}.
			}
		}
		\label{SuppFig3}
	\end{figure}
	%
	%\section*{Supplementary Video 1:}
	%\section*{Supplementary Video 2:}
	%\section*{Supplementary Video 3:}
\end{document}